\documentclass[twocolumn,apl,aps,superbib,tightenlines,floatfix]{revtex4}
\usepackage{amsfonts}
\usepackage{amsmath}
\usepackage{amssymb}
\usepackage{graphicx}

\begin{document}

\title{Etching-dependent reproducible memory switching in vertical SiO$_{2}$ structures}
\author{Jun Yao, Lin Zhong\cite{cross}, Douglas Natelson\cite{cross}, and James M. Tour\cite{cross}}
\affiliation{Departments of Applied Physics, Electrical \&
Computer Engineering, Physics \& Astronomy, Chemistry,  and
Computer Science, Rice University, MS 222, 6100 Main Street,
Houston, Texas 77005}

\begin{abstract}
Vertical structures of SiO$_{2}$ sandwiched between a top tungsten
electrode and conducting non-metal substrate were fabricated by
dry and wet etching methods. Both structures exhibit similar
voltage-controlled memory behaviors, in which short voltage pulses
(1 $\mu$s) can switch the devices between high- and low-impedance
states.  Through the comparison of current-voltage characteristics in
structures made by different methods, filamentary conduction at
the etched oxide edges is most consistent with the results,
providing insights into similar behaviors in metal/SiO/metal
systems. High ON/OFF ratios of over 10$^{4}$ were demonstrated.

%\medskip PACS Numbers: {72.25.Dc,73.23.Ad, 85.75.Nn}
\end{abstract}

%\keywords{}
\maketitle

Resistive switching materials have been studied intensively as
candidates for non-volatile memories. A
metal/insulator/metal (MIM) sandwich structure is usually adopted,
with "M" extending to good conducting non-metals\cite{1}. Many
oxides such as TiO$_{2}$ \cite{2}, NiO \cite{3}, Nb$_{2}$O$_{5}$
\cite{4}, SrTiO$_{3}$ \cite{5}, and perovskites \cite{6} have been
investigated and shown to have voltage- or current-controllable
bistable low-impedance (ON) and high-impedance (OFF) states. One
oxide, SiO$_{2}$, can work as a solid electrolyte in a metal-doped
switching system \cite{7}. Meanwhile, the amorphous form of SiO
can exhibit memory phenomena \cite{8,9,10,11,12,13}. In
such a system, several tens or hundreds of
nanometers of SiO is evaporated between two metal electrodes.
After electroforming, the voltage-controllable
conductance states can be varied by a factor up to 10$^{3}$.
Various models have been proposed to account for the
switching behaviors,
such as ion injection\cite{9},
a filamentary model \cite{10}, and defect introduction by
vacancies \cite{11}.

In this letter, we report reproducible resistive switching behaviors
in M/SiO$_{2}$/M vertical structures, fabricated by either wet or dry
etching methods.  We find that the switching in our system apparently
happens at the vertical edges produced by the etching process.  This
constrained conduction path, dependent on the etching methods, can
lead to an ON/OFF ratio exceeding 10$^{4}$, an order of
magnitude larger than that achievable in M/SiO/M systems \cite{8}.
The particular current-voltage (I-V) characteristics in structures
produced by different etching methods provide further insights into
the underlying filamentary character of conduction in SiO$_{x}$
switching systems.

Structures with two different oxide growth methods were examined.
SiO$_{2}$ with a thickness of 50 nm was grown by (1) thermal
oxidization of a silicon substrate and (2) plasma-enhanced
chemical vapor deposition (PECVD) on a TiN/Si substrate (TiN was
deposited by physical vapor deposition (PVD) on Si). Likewise,
vertical M/SiO$_{2}$/M structures were fabricated by two different
approaches. In the first approach, a lift-off process was used to
define circular tungsten (W) electrodes (by sputtering, with 5
nm-thick Ti adhesion layer) having thicknesses of 100 nm and
diameters of 50 $\mu$m. 10:1 buffered oxide etch (BOE, J. T.
Baker) was used to remove the surrounding SiO$_{2}$, leaving the
layer underneath the W electrode protected (device type 1 or
DEV-1, see inset in Fig. 1a). In the second approach, reactive ion
etching (RIE) was used to define the vertical sandwich structures.
10 nm of TiN and 100 nm of W were first deposited on SiO$_{2}$ by
PVD. A 70$\times$70 $\mu$m$^{2}$ photoresist area was then
patterned by photolithography and used as the sacrificial mask.
Corresponding etching recipes (e.g., SF$_{6}$/BCl$_{3}$/Cl$_{2}$
for W etching; BCl$_{3}$/Cl$_{2}$ for TiN etching; and
CF$_{4}$/CHF$_{3}$ for SiO$_{2}$ etching) were used with the
layered structure underneath the photoresist protected, thereby
forming the vertical structure (DEV-2, see inset in Fig. 2a). A
several-minute annealing at 600 $^{o}$C in an Ar/H$_{2}$
environment was performed before electrical characterizations.
Measurements were done using an Agilent 4155C semiconductor
parameter analyzer under a single sweep mode. Bias voltage was
applied by a probe tip at the top W electrode with the conducting
substrate grounded. All data was collected in vacuum (~10$^{-2}$
mTorr) at room temperature, unless otherwise specified.

\begin{figure}[tbp]
\includegraphics*[width=8.5cm]{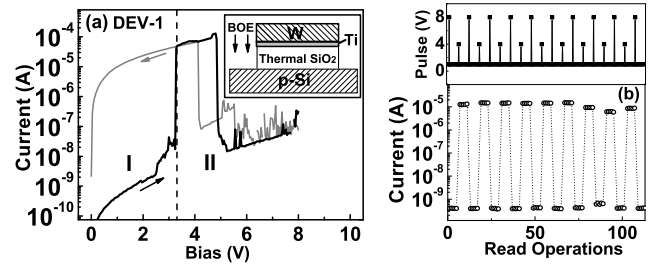}
\caption{(Color online) (a). I-Vs of a forward (0$\rightarrow$8 V)
and subsequent backward (8$\rightarrow$0 V) sweeps in DEV-1. The
inset is a schematic of the device. The dashed line divides the
bias regions of "I" and "II".  (b). A series of reading the device
state by applying bias pulses of +1 V (1 $\mu$s). After every five
readings, the device was set by an erasing pulse (+8 V, 1 $\mu$s)
or a writing pulse (+4 V, 1 $\mu$s) as show in the corresponding
upper panel.}
\end{figure}

Fig. 1a shows the typical I-Vs in a formed DEV-1 device under a forward
(0$\rightarrow$8 V) sweep and then a backward (8$\rightarrow$0 V)
sweep of the top electrode relative to the bottom one. The forward
sweep features a high-impedance state at low bias
(region "I" in Fig.
1a), with a sudden current jump at $\sim$3.3 V and then a sudden drop
at $\sim$5 V back into a high-impedance state. The backward sweep,
from high bias back toward 0 V, shows a low-impedance state at
region "I".  This hysteretic behavior results from a
voltage-controlled resistance change in the higher bias
(writing) range (region "II" in Fig. 1a), indicated by
the sudden rise in current during
the forward sweep. A rapid falling edge of voltage in this region
can write the device into a conductance state corresponding to
this ending voltage\cite{8}. For example, a rapid voltage drop at
4 V writes the device into an ON state, while one at 8 V erases
the device into OFF
%(In Fig. 1a, the forward sweep was obtained
%after at least one forward sweep performed to the same bias range;
%the device had been in an OFF state since a very rapid voltage
%drop to 0 V, at the end of the previous forward sweep, is expected
%to place the device into the corresponding high-impedance state).
The obtained state can be read out at the low bias region "I" without
being destroyed, allowing the structure to act as a non-volatile
memory device.  Pulses as narrow as 1 $\mu$s (limited by
instrumentation) were applied to read, write and erase the device
(Fig. 1b), with an ON/OFF ratio over 10$^{4}$ achieved. The memory
states are stable against reading, showing no degradation in ON- or
OFF-state conduction after 1000 read operations.

\begin{figure}[tbp]
\includegraphics*[width=8.5cm]{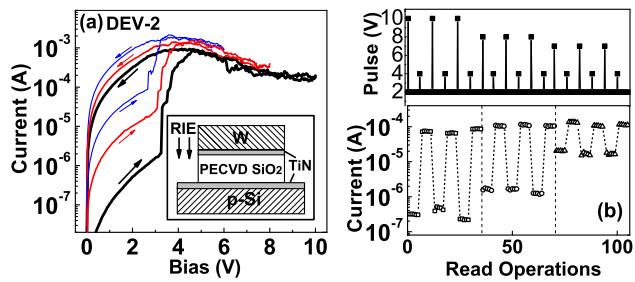}
\caption{(Color online) (a). A set of I-Vs by forward and
subsequent backward sweeps in DEV-2, with black (bottom), red
(middle), and blue (top) curves corresponding to sweep bias ranges
of 0-10, 0-8, and 0-6 V, respectively. Inset is a device
schematic. (b). A series of reading the device state by applying
bias pulses (1 $\mu$s) of +2 V (five reads between each set of
writing-erasing pulses). The device undergoes OFF-state changes by
applying erasing pulses (1 $\mu$s) of different magnitudes of +10
V, +8 V, and +7 V (with the writing pulse +4 V unchanged) as shown
correspondingly in the upper panel.}
\end{figure}

Similarly, Fig. 2a (black curves) shows the typical hysteresis I-Vs
in DEV-2 devices with bias sweeps as described
above. Compared to those in Fig. 1a from the DEV-1 device,
the I-Vs here have (1) higher ON and OFF currents and (2)
comparatively smooth current changes in the writing region. The
conductance change in this region, without a drastic
rise or drop, means that the conductance state of the device can
be changed semi-continuously by applying bias pulses of different
amplitudes. The color curves in Fig. 2a show how different bias
sweep ranges (thus different final writing voltages)
change the conduction states of the device
%(again,
%each set of forward and backward sweeps was obtained after at
%least one forward sweep preformed to the same bias range).
Decreasing of the sweep range leads to a gradual decline in
the ON/OFF ratio because of an increasing current in the OFF state
set by the reduced final bias (the ON current tends to increase
at a rate much smaller than that of increase in the OFF current).
The writing region tends to shift toward low bias,
indicated by the shift of the current-rise edge in the forward
sweep. A multilevel or analog memory \cite{8,14} is demonstrated
in Fig. 2b by applying erasing bias pulses of different
amplitudes. The adjustable ON/OFF ratio is less than
10$^{3}$ due to comparatively large OFF currents.

%%%

To clarify where the switching takes place in these vertical devices,
W or W/TiN electrodes with the same thicknesses and sizes were
deposited on the (PECVD) SiO$_{2}$/TiN/Si substrate, without doing any
etching of the oxide (see left schematic in Fig. 3a), or on
wet-etching defined SiO2 pillars of larger diameter on the (thermal)
SiO$_{2}$/Si substrate (see right schematic in Fig. 3a).  The samples
were annealed under the same conditions as those adopted for the
previous structures and then characterized via electrical
measurements.  No conduction was observed up to a bias of 25 V
(I$\sim$10$^{-12}$ A).  Devices with different diameters (25 $\mu$m,
50 $\mu$m, and 100 $\mu$m) were also made and the ON currents were
collected for DEVs-1 and another type of wet-etching defined
W/Ti/(PECVD) SiO$_{2}$/TiN/Si devices (DEVs-3, see schematic in
Fig. 3b).  For both DEVs-1 and DEVs-3, the ON currents scale
approximately with the diameter (black dashed lines in Fig. 3d-e)
rather than with the device area (red dotted lines in Fig. 3d-e).
These results imply that conduction and switching only take place
after the etching and are likely localized at the vertical SiO2 edges.
The I-Vs of DEVs-3 tend to have features combining attributes of DEV-1
and DEV-2.  DEVs-3 typically have higher ON and OFF currents than
those in DEV-1, but lower than those in DEV-2 (at the same bias sweep
ranges).  The forward sweep still begins with a sudden current rise in
the writing region, but then follows a less intense declining
tail. Fig. 3c shows three forward I-Vs from each of the three types of
devices.  We suspect that surface differences resulting from etching
methods and oxide growth techniques are responsible for this
variation.
%One indicator of structural differences between oxides
%grown by varying methods is the difference in wet etching rates of
%PECVD SiO$_{2}$ and thermal SiO$_{2}$.

\begin{figure}[tbp]
\includegraphics*[width=8.5cm]{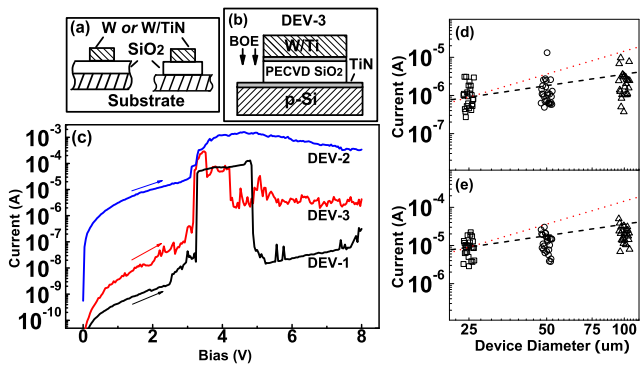}
\caption{(Color online) (a). Schematics of devices with (left) no
SiO$_{2}$ etching and (right) etched SiO$_{2}$ pillar with a
smaller on-top electrode. (b). Schematic of DEV-3. (c). A set of
forward sweeps starting with the OFF states in the formed vertical
structures of DEV-1 (black curve), DEV-2 (blue curve), and DEV-3
(red curve). (d). ON currents at +1 V (written by 1 $\mu$s, +5V
pulses) for DEVs-1 and (e) ON currents at +1 V (written by 1
$\mu$s, +4 V pulses) for DEVs-3, with each type having device
diameters of 25 $\mu$m (rectangles), 50 $\mu$m (circles), and 100
$\mu$m (triangles). 25 data points/devices were collected for each
size. The dashed black lines describe a (diameter) scaling and the
dotted red lines scaling.}
\end{figure}

The surface nature of the conduction is further supported by the response of
devices to annealing in a reducing atmosphere. A several-minute
thermal annealing at 600 $^{o}$C in Ar/H$_{2}$ was necessary to
observe the switching in wet etched devices.
Before annealing, a majority (over 80\%) of the devices was
non-conducting (e.g., I$\sim$10$^{-12}$ A at a bias up to 25V).
Detectable conduction began to take place after the thermal
annealing in the reducing environment.  An electroforming
process takes place by sweeping to high
voltages (e.g., 20 V).   Large current fluctuations gradually move
toward lower bias voltages in subsequent sweeps. Finally, reproducible
forward I-Vs as described previously are established. For the devices
fabricated by dry etching (DEV-2), a similar forming process can
take place even before the annealing.  An annealing simply obviates
this forming process by introducing the device into a formed ON
state for the first forward sweep.

All three device types have similarities once operational. They
have (1) similar reading/writing bias ranges (see Fig. 3c), (2)
similar noise distributions with comparatively smooth I-Vs at the
reading bias range and larger fluctuations in the writing region,
and (3) fast switching time down to 1 $\mu$s. Low-temperature
tests show that these devices cannot be electroformed at a
temperature below 150 K, while the ON-state conductance shows
relatively little temperature dependence down to 100 K \cite{15}.
Our other tests show that the switching and forming processes in
all types of the devices are bias-polarity independent. Since all
the devices are asymmetric in structure with non-metallic
substrates of either Si or TiN/Si, metal filament formation is
unlikely to be responsible for the observed switching because such
a process usually involves metal ion migration or injection from
the electrode, which is bias-polarity dependent \cite{1}.
Moreover, metal filaments generally have higher ON currents than
those observed here.  All the observed characteristics resemble
those in the M/SiO/M system \cite{8}, where SiO plays the
essential switching role. In our structures we suggest that
nonstoichiometric SiO$_{x}$ at the etched oxide edges is the
switching medium.

While M/SiO/M systems were first studied
decades ago, the detailed underlying switching mechanism remains
debatable \cite{8}.
%The layered structure of
%SiO in the M/SiO/M system along with comparatively large and
%smooth currents contributed from the entire SiO layer makes it
%challenging to discern whether the switching is a bulk effect or
%is localized.
The surface-restricted switching in our system may offer insights.
The I-Vs in DEVs 1-3 mainly differ in the
current magnitudes and fluctuations.  As shown in Fig. 3c, with
the decreasing of the operating current level, the current
fluctuations in the writing bias region increase.
%(consider, for
%example, the large rises and drop in currents as the largest
%fluctuations)
Note that in Fig. 3c the OFF current of DEV-2 is
much larger than that of DEV-1, changing from a relatively smooth
I-V into one with discrete steps.  We suggest that this is a
consequence of having an ensemble of conducting paths in DEV-2 and
only a few paths in DEV-1.  The sweep-to-sweep variation in switching voltages
even in the same DEV-1 further indicates the
discrete nature of conduction paths.  One can imagine that a large
ensemble of DEVs-1 in parallel would lead to the
comparatively smooth I-V in DEV-2 due to a variety of current rise
and drop edges. The I-V of DEV-3 in Fig. 3c is consistent with
this idea, with an intermediate conductance device DEV-3 having
features between the two limits of DEV-1 and DEV-2.
%These features
%further support discrete-path or filamentary conduction in our
%system at nonstoichiometric device edges, in accordance with
%expectations based on studies of M/SiO/M devices by different
%methods [12, 13].
It is straightforward that a
large number of paths with non-uniform writing/erasing
biases would limit the overall ON/OFF ratio, while reducing the path
number can push this ratio up, as demonstrated in DEV-1 with an
ON/OFF ratio $> 10^{4}$.
Currently the device size is limited due to our instrumentation
(e.g., the size of probe tip for measurement).  The variation in ON
currents for devices with the same nominal sizes (see Fig. 3d-e)
implies that the switching likely takes place locally at some
parts of the SiO$_{2}$ vertical edge instead of uniformly along
the entire circumference.  It is thus expected that the device size
could be further reduced, even down to one comparable to the
actual switching region, which could be small due to the
filamentary nature.  Devices can have OFF
currents at the noise level of our instrumentation ($\sim$
10$^{-12}$ A), giving an ON/OFF ratio approaching 10$^{6}$
\cite{15} even at the current device size. These traits indicate
the possibility of high-density memory arrays based on this
switching.

%We have demonstrated reproducible resistive switching with memory
%properties in vertical sandwich structures with SiO$_{2}$ as the
%switching layer. The electrical characteristics depend
%significantly on the oxide etching method, as well as annealing
%history in reducing environment and oxide growth method, currents
%scale with device perimeter rather than area. These traits suggest
%that the switching is filamentary in nature and takes place in the
%stoichiometrically poor edges of the etched oxide. The localized
%character of conducting paths and the critical role of etched
%oxide surfaces raise the possibility of nano-scale SiO$_{x}$ based
%memories, though an improved understanding of the detailed
%switching mechanism would be essential. The question of the
%formation of individual conduction path remains open. The
%heat-assisted electroforming processes in our devices and the
%non-working state at low temperature may indicate a
%fusion-and-reforming mechanism induced by local heating due to hot
%electrons.

JY and JMT acknowledge partial support from an Army Research
Office SBIR through a subcontract with Integrated Sensor Solutions
(now Privatran). DN acknowledges the support of the David and
Lucille Packard Foundation. LZ acknowledges the support from the
Texas Instruments Leadership University Fund and National Science
Foundation Award \#0720825.

\clearpage

{\center{\bf Supplementary material for ``Etching-dependent reproducible memory switching in vertical SiO$_{2}$ structures''}}

\renewcommand{\thefigure}{S\arabic{figure}}
\setcounter{figure}{0}

Due to space limitations in the main manuscript, here we show two supplemental figures that provide further information about the experiments.

\begin{figure}[ht]
\begin{center}
\includegraphics[clip, width=15cm]{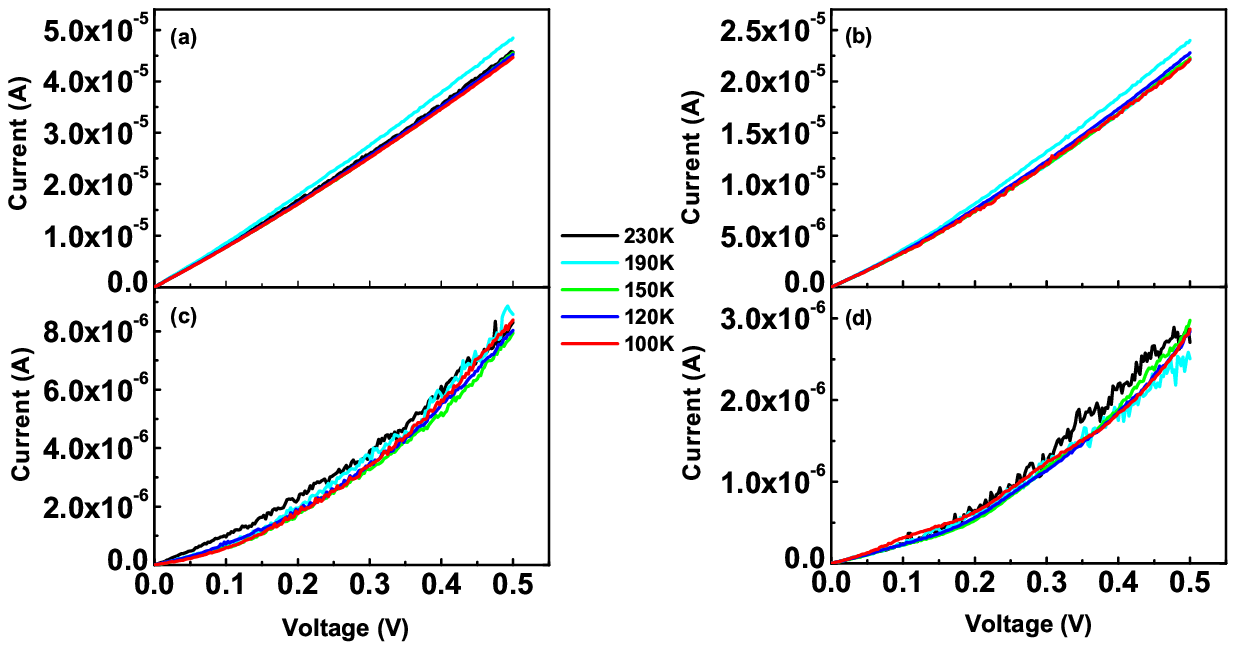}
\end{center}
\caption{
$I-V$s measured at different temperatures for four individual DEVs-3 (D = 50~$\mu$m). The devices were set at different conducting states, from high to low (a$\rightarrow$d). They indicate little temperature dependence for the conduction down to 100~K. 
}
\end{figure}

\begin{figure}[ht]
\begin{center}
\includegraphics[clip, width=8.5cm]{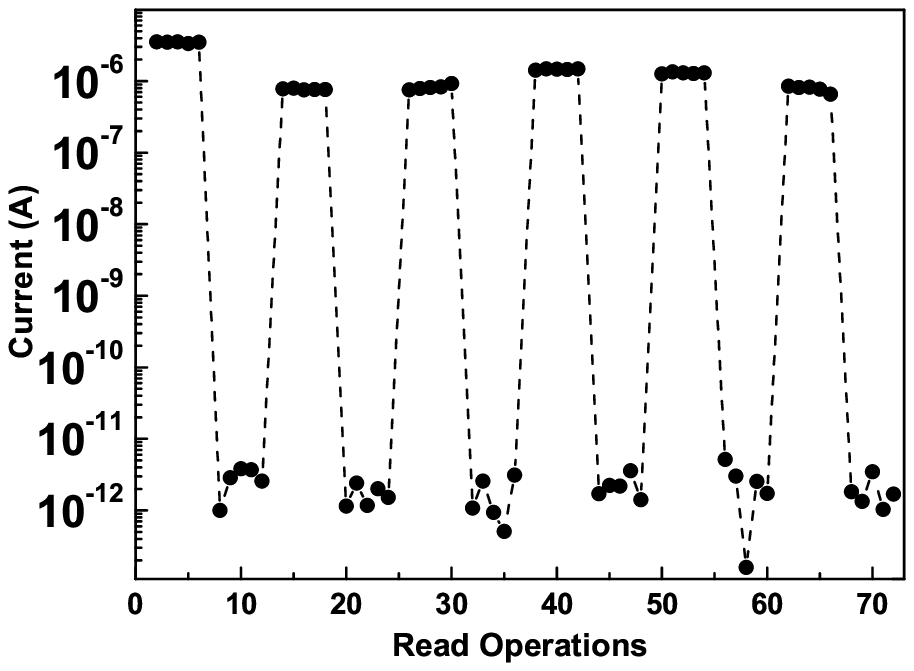}
\end{center}
\caption{
Memory-state readings of a DEV-1 ($D = 25~\mu$m) by applying bias pulses of +1~V (1~$\mu$s). The device was either written into an ON state by a (+5~V, 1~$\mu$s) pulse or erased into an OFF state by a (+10~V, 1~$\mu$s) pulse after every five readings. It shows that the OFF currents are at the noise level of our instrumentation (10$^{-12}$~A) and an ON/OFF ratio approaching to $10^6$.
}
\end{figure}

\end{document}